\documentclass[aps,superscriptaddress,eqsecnum,nofootinbib,preprintnumbers]{revtex4}
\usepackage{graphicx,epsfig,axodraw}
\usepackage{amsmath}
\usepackage {amssymb}

\newcommand{\be}{\begin{eqnarray}}
\newcommand{\ee}{\end{eqnarray}}
\newcommand{\bea}{\begin{eqnarray}}
\newcommand{\nn}{\nonumber}
\newcommand{\eea}{\end{eqnarray}}

\newcommand{\reef}[1]{(\ref{#1})}

\def\de{\partial}
\def\a{\alpha}
\def\b{\beta}

\def\d{\delta}

\def\e{\eta}
\def\la{\lambda}
\def\La{\Lambda}
\def\k{\kappa}
\def\m{\mu}
\def\n{\nu}
\def\r{\rho}
\def\o{\omega}

\def\f{\phi}

\def\ep{\epsilon}
\def\th{\theta}

\def\C{{\cal{C}}}

\def\ie{{\it i.e. }}

\begin{document}

\title{The consistency of codimension-2 braneworlds and their cosmology}
\author{Christos Charmousis}\email{christos.charmousis@th.u-psud.fr}
\affiliation{LPT, Universit\'e de Paris-Sud, B\^at. 210, 91405 Orsay
CEDEX, France} \affiliation{Laboratoire de Math\'ematiques et
Physique Th\'eorique (LMPT) CNRS : UMR 6083, Universit\'e Fran\c
cois Rabelais - Tours}

\author{Georgios Kofinas}\email{gkofin@phys.uoa.gr}
\affiliation{Department of Physics and Institute of Plasma Physics,
 University of Crete, 71003 Heraklion, GREECE}

\author{Antonios Papazoglou}\email{antonios.papazoglou@port.ac.uk}
\affiliation{Institute of Cosmology and Gravitation, University of Portsmouth, Portsmouth PO1 3FX, UK}

\date{\today}

\begin{abstract}

We study axially symmetric codimension-2 cosmology for a
distributional braneworld fueled by a localised four-dimensional
perfect fluid, in a six-dimensional Lovelock theory.  We argue that
only the matching conditions (dubbed topological) where the
extrinsic curvature on the brane has no jump describe a pure
codimension-2 brane. If there is discontinuity in the extrinsic
curvature on the brane, this induces inevitably codimension-1
distributional terms. We study these topological matching
conditions, together with constraints from the bulk equations
evaluated at the brane position, for two cases of regularisation of
the codimension-2 defect. First, for an arbitrary smooth
regularisation of the defect and second for a ring regularisation
which has a cusp in the angular part of the metric. For a
cosmological ansatz, we see that in the first case the coupled
system is not closed and requires input from the bulk equations away
from the brane. The relevant bulk function, which is a
time-dependent angular deficit, describes the energy exchange between the brane and
the six-dimensional bulk spacetime. On the other hand, for the ring
regularisation case, the system is closed and there is no leakage of
energy in the bulk.  We demonstrate that the full set of matching
conditions and field equations evaluated at the brane position are
consistent,  correcting some previous claim in the literature which
used rather restrictive assumptions for the form of geometrical
quantities close to the codimension-2 brane. We analyse the modified
Friedmann equation and we see that there are certain corrections coming from
the non-zero extrinsic curvature on the brane. We establish the presence of
geometric self-acceleration and a
possible curvature domination wedged in between the period of matter
and self-acceleration eras as signatures of codimension-2 cosmology.

\end{abstract}

\maketitle

\section{Introduction}

Codimension-2 distributional defects{\footnote{Codimension denotes
the number of independent normal vectors to the worldsheet swept by
the defect. We concentrate on the codimension since it is this,
rather than the number of tangential dimensions of the defect, that
essentially quantifies its dynamical properties.}}  are
infinitesimally thin matter filaments which are particularly
interesting, subtle and elusive objects in gravitational theories.
In four-dimensional general relativity, codimension-2 defects of
infinitesimal size describe cosmic strings which are the singular
relativistic versions of finite size vortices in condensed matter
systems. Strings were thought to play a dominant role in the
structure formation history of our Universe \cite{kibble}. They were
widely studied in differing contexts, but mostly at the zero
thickness limit and in the test approximation;  in other words,
neglecting their self-gravity, \ie the effect of their
energy-momentum tensor on their proper evolution. When one takes
into account their self-gravity, at least  for a straight
distributional string, one finds that they are described by a
codimension-2 conical singularity \cite{vilenkin} sourcing the
constant tension of the defect (see also \cite{linet} for constant
curvature bulk). Their effect is a global, topological one, since
locally the induced gravitational field is flat. Indeed, each point
of the string worldsheet generates a two dimensional cone with a
non-trivial deficit angle.

When considering however, non-trivial geometries, one was confronted
by a paradox \cite{israel} which can be summed up in the following
way: test cosmic strings are described using the Nambu effective
action which leads to a local minimal area requirement,  or
equivalently a minimal surface spanned by the string worldsheet.
This is geometrically described by the trace of the second
fundamental form  being equal to zero (for examples see
\cite{bonjour}) and can be visualised as the surface (with boundary)
of a soap film obtained upon retrieving slowly a circular wire from
liquid soap.  To the Nambu action one can find specific extrinsic
and intrinsic curvature corrections \cite{Ruth} in order to take
into account the finite width of the defect. It is natural therefore
to expect, that upon considering the {\it{self}}-gravitating field
of the cosmic string, a small correction should result on top of the
minimal surface motion. Surprisingly however, self-gravity of the
distributional string leads to trivial motion or non-bending of the
worldsheet with the {\it {full}} second fundamental form being
identically zero. Geometrically the string worldsheet is a totally
geodesic surface. In other words self-gravity makes the defect
completely rigid not allowing it to bend in the ambient spacetime
\cite{israel2} giving back again essentially the straight string
metric (see also \cite{cbl}).

This puzzling  property of gravitational rigidity reemerged in the
context of codimension-2 braneworlds embedded in a six-dimensional
spacetime. Unlike four-dimensional General Relativity (GR), where one could convincingly
argue that  self-gravity of cosmic strings is negligible, this was
far more subtle for a braneworld given that its motion is by
construction fueled by its proper matter density. Indeed, not
surprisingly, early work on codimension-2 distributional braneworlds
showed that they were pure tension objects \cite{klein}. In other
words, one could not find a distributional solution (see \cite{gt} for a mathematical
discussion on distributional sources in GR) for a non
constant energy-momentum tensor such as that, say, of a perfect
fluid necessary to describe braneworld cosmology for example. One
had to introduce finite thickness effects \cite{thick}, which made the problem
considerably more difficult \cite{cosmothick}, and more importantly, plagued the
generality of the result and hence its physical relevance. Indeed,
the key point when studying distributional sources is that they
describe the important main features of brane dynamics for an
arbitrary family of finite width regularisations at the limit of
infinitesimal thickness. At the absence of a distributional
description the dynamics are reguralisation dependent. The
distributional description for example, leads to considerable
simplification when one looks at codimension-one braneworlds where
one can solve the full system of brane-bulk equations at least for
cases with enhanced symmetry such as homogeneous and isotropic
cosmology \cite{bow}.

Activity on this subject concentrated mainly on the interesting
topological properties of codimension-2 defects in relation to the
self-tuning paradigm \cite{brucelee}. The key point to resolve the
gravitational rigidity puzzle was to understand that it was not the
defect construction which was problematic, rather the gravity theory
itself did not have the relevant differential complexity in order to
describe complicated distributional solutions{\footnote{We thank
Brandon Carter for early, enlightening discussions on the subject of
gravitating distributional defects.}}. In other words, it is not
strings that cannot bend, it is just Einstein equations that cannot
describe the curving of strings infinitesimally. Even ordinary
codimension-1 junction conditions fail in cases where the
spacetime is of lesser symmetry, such as when pasting together the
Kerr metric with flat spacetime, where the spacetime metric fails to
be continuous.

In fact, given that the defect carries a distributional
energy-momentum tensor, the gravitational field equations have to be
second order in such a way as to have piecewise continuous first
metric derivatives. Although four-dimensional GR is the unique second
order tensor theory with this property, in five or six dimensions one has
to add an extra term, the Gauss-Bonnet term, to the action in order
to have the most general second derivative (and not higher than
second derivative) field equations. The generic second order
derivative tensor gravity theory is in fact well-known to be
Lovelock's theory \cite{Lovelock} (for a review see \cite{deruelle},
\cite{charmousis}).  In this gravity theory each $2n+1$ dimensional
manifold picks up in the gravitational action a novel Langrangian
density, which is  a specific combination of the $n$th power of
spacetime curvature. This gravitational term stems from the
Euler-Poincar\'e characteristic of the manifold in $2n$ dimensions
and hence is purely topological at this dimension.  As such, the
Einstein-Hilbert and the Gauss-Bonnet terms are the Euler-Poincar\'e
characteristics for two-dimensional and four-dimensional manifolds
without boundary.

The clear-cut hint that the completion of GR in higher dimensions
may be relevant to codimension-2 braneworlds came with the work of
Bostock et al. \cite{bos}, where it was noticed that upon
considering the general second derivative gravity theory, one could
have, at least  in principle, a non trivial energy momentum tensor
fueling geometric junction conditions for a codimension-2 conical
defect. Furthermore, it was shown that the higher order gravity term
in question, the Gauss-Bonnet term, was generating on the brane an
induced Einstein-Hilbert term plus extrinsic curvature corrections,
whereas the Einstein-Hilbert term in the bulk was giving a pure induced
cosmological constant. This observation was urged further in \cite{cz} (see also \cite{KS}) where it was shown that
Lovelock theory can source defects with codimension even higher than two, unlike
ordinary GR.  A simple geometric explanation was given relating by a
simple sum rule each $n$th bulk Lovelock density with the distributional
brane's even codimension $2m$ and the induced brane Lovelock density,
$n-m$. Therefore, the six-dimensional bulk Gauss-Bonnet term ($n=2$) is
"reduced" to a codimension-2 Einstein-Hilbert term on the brane $n-m=1$
and so forth. The same $n=2$ Gauss-Bonnet term in eight dimensions would
"reduce" to a pure tension term on a codimension-4 brane $n-m=0$ and so
on.

This  activity came to a halt when under some quite mild symmetry
hypothesis (axial symmetry) it was claimed that the full set of
junction plus bulk field equations at the location of the brane led
to an inconsistent system of differential equations for a
non-trivial distributional codimension-2 defect \cite{thereof}.
Furthermore, the junction conditions were shown not to be unique
\cite{cz}, without any obvious particular physical difference but with only
differing mathematical regularity. Lastly, certain restrictive initial conditions
on the braneworld surface \cite{bos} introduced important
constraints \cite{pap2} on the admissible matter on the brane.

The aim of this paper is to falsify or explain all points in the
previous paragraph. We will show that the system of equations is not
only consistent, but has in general a free degree of freedom in the face of a
varying deficit angle function. This will result from carefully
considering the correct and all relevant geometric expansions for
the braneworld geometry. We will show that differing mathematical
regularity leading to differing junction conditions inevitably leads
to {\it distinct} physical setups (following up work by \cite{KS}). We shall show that the unique junction conditions
tailored for a pure codimension-2 defect have the {\it same}
mathematical regularity as those of GR and are the topological
matching conditions of \cite{cz}. The matching conditions introduced in
\cite{bos} inevitably lead to extra codimension-1 distributional
matter. Having established these basic facts, we will go on to show
that one can obtain the modified Lema\^itre-Friedmann-Robertson-Walker (LFRW) brane equations by solving the
full system of the field equations, as well as the junction
conditions in the brane neighborhood. We will find that the
cosmological equations, apart from the ordinary LFRW term,
have several interesting features including a geometric
self-acceleration (in agreement with the maximally symmetric solutions of \cite{papapapa}), brane bending effects depending on the equation
of state of matter, and energy exchange between the bulk and the
brane triggered by a dynamical conical deficit angle.

The organisation of the paper is as follows: we will give the general
set-up and a careful treatment of the differing matching conditions
(topological, topological with ring regularisation and general) in
the next section. We will then demonstrate consistency of the whole
set-up and go on to investigate the cosmological equations. We will
conclude in the last section.

\section{General setup and matching conditions}

\par
Let us consider the general system of six-dimensional Lovelock gravity coupled to
localised brane sources. If we describe the regular bulk matter by  $S_{bulk}[\verb"g"]$ and the
 distributional brane  matter  by $S_{brane}[g]$, the total action of the system is,
 \be
\!\!\!\!\!\!\!S=\frac{1}{2\kappa_{6}^{2}}\!\int
\!d^6x\sqrt{\!-|\verb"g"|}\, \Big(\mathcal{R}+{\alpha \over 6}\,
\mathcal{L}_{GB}  \Big)\!+ \! S_{bulk}[\verb"g"]\!+ \!
S_{brane}[g]~, \label{6action}\ee with the Gauss-Bonnet Lagrangian
density \be \mathcal{L}_{GB} = \mathcal{R}^{2}\!-4\mathcal{R}_{MN}\,
\mathcal{R}^{MN}+\mathcal{R}_{MNK\La}\,\mathcal{R}^{MNK\La}~. \ee In
the above, the calligraphic quantities refer to the bulk metric
tensor $\verb"g"$, while the regular ones to the brane metric tensor
$g$. The field equations arising from the action (\ref{6action}) are
\be \mathcal{G}_{M}^N\!-\!{\alpha \over 6}\mathcal{H}_{M}^N
=\kappa_{6}^2 \mathcal{T}_{M}^N+\kappa_{6}^2\,T_{M}^N~,\label{6eqs}
\ee where $\mathcal{T}_{MN}$ is a regular bulk energy-momentum
tensor and  $T_{MN}$ is the distributional  brane energy-momentum
tensor. The most plausible assumption is that the bulk contains only
cosmological constant $\mathcal{T}_{MN}=-(\Lambda_{6}/\k_6^2)
~\verb"g"_{MN}$, however, for the sake of generality we leave for
the moment $\mathcal{T}_{AB}$ arbitrary and non-vanishing (we will
however assume later that it is regular at the position of the
brane). The Gauss-Bonnet contribution to the equations of motion is
explicitly \be \mathcal{H}_{MN}\equiv \!\frac{1}{2}\mathcal{L}_{GB}~
\verb"g"_{MN}\!-\!2\mathcal{R}\mathcal{R}_{MN}
\!+\!4\mathcal{R}_{MK}\mathcal{R}_{N}^{\,\,\,\,K}\!+\!
4\mathcal{R}_{MKN\La}\mathcal{R}^{K\La}\!-\!2\mathcal{R}_{MK\La
\Xi}\mathcal{R}_{N}^{\,\,\,\,K\La \Xi}~. \ee In the following, we
will fix the notation to $3 \kappa_{6}^2 = 4 \pi$ to simplify the
equations.

Let us now consider that there is axial symmetry in the bulk, so
that the bulk metric ansatz can be written in the brane
Gaussian-Normal coordinates as \bea
ds_{6}^2=dr^2+L^2(x,r)d\th^2+g_{\mu\nu}(x,r)dx^{\mu}dx^{\nu}
=dr^2+g_{ab}(x,r)dx^adx^b ~,\label{6metric} \eea where $\th$ has the
standard periodicity $2\pi$, and the braneworld metric
$g_{\mu\nu}(x,0)$ is assumed to be regular everywhere, with the
possible exception of isolated singular points on the brane. We
denote by $y$ collectively the transverse coordinates $r,\theta$.

To study in detail the matching conditions we must first differentiate between two distinct cases:
\begin{enumerate}

\item The first case is {\it topological matching conditions}, discussed in \cite{cz}, which have a geometric origin based on the distributional version of the Chern-Gauss-Bonnet theorem \cite{zz}. They assume everywhere smooth intrinsic and extrinsic tangential sections.

\item The second case, which admits the lesser mathematical regularity, was introduced in \cite{bos} and assumes, not only a conical deficit based on the normal geometry, but also a distributional jump in the extrinsic tangential sector, namely a specific combination of extrinsic curvature quantities\footnote{In \cite{bos} a particular jump of the extrinsic curvature was assumed, namely that the extrinsic curvature vanishes at the core of the defect.}. We will call this possibility the {\it general case} since for regular extrinsic parts it reduces to the topological case outlined above.

\end{enumerate}

As we shall see later on, the second type of matching conditions
inevitably lead to additional codimension-1 distributional
singularities \cite{KS}, and hence additional matter sectors {\it
have} to be introduced in order to close the system mathematically.
This is not too surprising, since mathematical discontinuities in the
extrinsic sector are expected to produce codimension-1 defects
\cite{israel}, \cite{davis}. On the other hand, the topological type
of matching conditions lead to regular induced metrics unlike the
former case.

\subsection{Topological matching conditions: Smooth regularisation}

Let us start with discussing the case of the {\it topological
matching conditions}. In this case, there is no jump of the
extrinsic curvature inside and outside the thin defect.  As usually
done when dealing with distributional sources, the matching
conditions are derived by integrating around the singular space.
Here, we seek to integrate over the conical space, say $\C$, with
internal metric \bea
&&ds_2^2=dr^2+L^2 d\theta^2~,\\
&&{\rm with} \ \ \  L(x,r)\!= \beta(x)r + {1 \over 2}\b_2(x) r^2 +
{1 \over 6}\b_3(x) r^3 +\mathcal{O}(r^{4})~, \label{Lexp} \eea where
$\d=2\pi (1-\beta(x))$ is the conical deficit. Notice, the
coefficients are allowed to depend on the  ``brane" coordinates
$x^{\m}$. The angle $\theta$ varies in the interval $[0,2\pi)$.

In the present paper, we are interested in the case that the
distributional brane energy-momentum tensor supports a codimension-2
Dirac singularity and therefore can be written as \be T_{MN} \equiv
T_{\m\n} {\d(r) \over 2 \pi L}\d_M^\m \d_N^\n ~. \label{co2emt} \ee
In general, there can be also codimension-1 $\d(r)$ parts in the
energy momentum tensor, which, as we will see later, is unavoidable
if the extrinsic curvature has a jump across the interface layer of the defect.

Then, by integrating the equations of motion, the only non-zero
contribution  yielding the codimension-2  matching conditions reads
\cite{cz},
 \be \a~{1 \over 4\pi}\int \!d^{2}y \sqrt{g_2}\,R_2 ~\left[ G_{\m\n} + W_{\m\n}- {3 \over 2\alpha} g_{\m\n}\right] = T_{\m\n} ~\label{1stjunc},
\ee where $R_2 = -2 L''/L$ is the curvature of the two-dimensional
internal space, and a prime denotes $\partial_{r}$\,. The Einstein
equation is corrected with the following contribution from the
extrinsic curvature $K_{\m\n} \equiv {1 \over 2}\de_r g_{\m\n}$  on
the brane \be W_{\m\n}= K_\m^\la K_{\n \la} - K K_{\m\n} + {1 \over
2} g_{\m\n} (K^2 - K_{\k\la}^2)~, \ee which is identical to the one
in \cite{bos}.

\begin{figure}[t]
\begin{center}
\epsfig{file=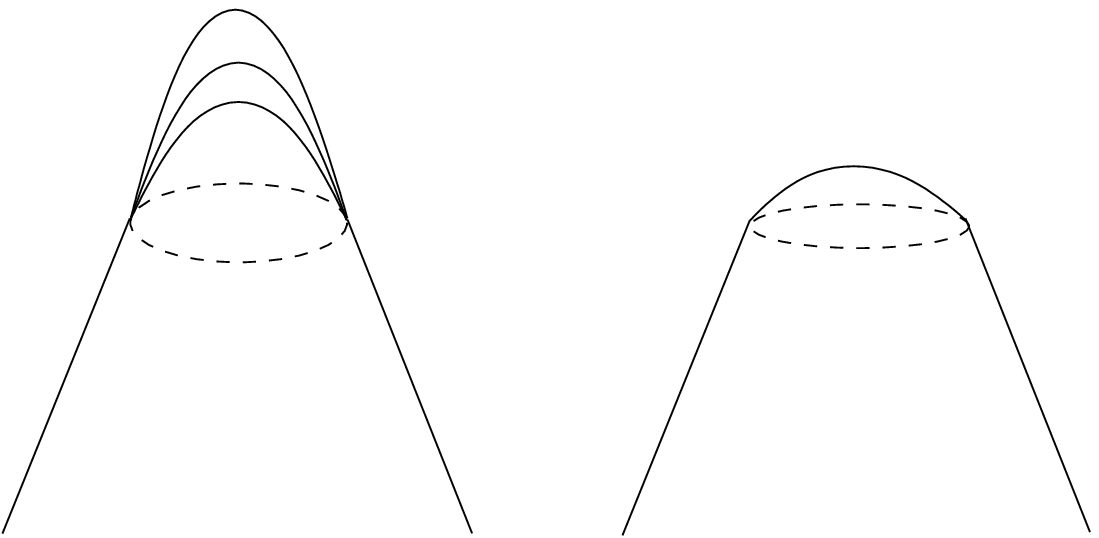,width=10cm,height=4cm}
\begin{picture}(50,50)(0,0)
\Text(-225,120)[c]{$C_\la$}
\Text(-225,50)[c]{$\de C$}
\Text(-32,82)[c]{$L^-$} \Text(-20,60)[c]{$L^+$}
\Text(-65,90)[c]{$r=0$}  \Text(-65,55)[c]{$r=\ep$}
\end{picture}
\end{center}
\caption{On the left, the conical singularity is regularised by a family of regular interiors $C_\la$ smoothly connected to the exterior solution of the cone at the boundary $\de C$. On the right, a ring regularisation is chosen where the internal space function $L$ is not smooth, but has a cusp across the ring at $r=\ep$.} \label{regularconelambda}
\end{figure}

In order to evaluate the integral of \reef{1stjunc} which is
singular at the tip, we assume a family of {\it smooth}
regularisations $C_\la$ of the cone, parametrised by $\la$,  with
smooth $C^2$ everywhere  caps (for details see \cite{cbl}), pasted
smoothly to the rest of the cone at a boundary $\de C$, which we
assume does not depend on the parameter $\la$ (see left part of
Fig.\ref{regularconelambda}). The distributional limit is obtained
at the $\lambda\rightarrow \la_c$ limit, where the cap tends to the
singular conical space. The width of the regularised defect is supposed to be small
enough so that the $r$-dependent brackets can be approximated with the leading
term in their $r$-expansion, which is just the value of the brackets at $r=0$.  Then the
integral has to be performed over only the internal curvature.
 The Chern-Gauss-Bonnet theorem relates the
geometric curvature for arbitrary $\lambda$, to the Euler
characteristic $\chi$ of the regularised cap and the integral of the
geodesic curvature $k_g$ of the boundary $\de C$. Then, for every
cap labeled by $\la$, the following holds \be {1 \over
4\pi}\int_{C_\la}\!d^{2}y \sqrt{g_\la} \,R_\la = \chi - {1 \over
2\pi }\int_{\de C_{\la}} \!k_g \,d\theta ~. \ee Since the
$\la$-labeled caps are topologically equivalent to a disk which has
$\chi=1$ and since $k_g=L'$, the theorem gives that for every kind
of smooth regularisation of the conical singularity, one obtains \be
{1 \over 4\pi}\int_{C_\la} \!d^{2}y\sqrt{g_\la} \,R_\la = 1 - \b ~.
\ee The above result only used the fact that the expansion of $L$
close to the conical defect has the expansion \reef{Lexp} and is
independent of the smooth regularisation of the interior of the
defect. Finally, with the use of the above theorem, the
codimension-2 matching conditions for the topological case is \be
G_{\m\n} + W_{\m\n}- {3 \over 2 \a}g_{\m\n} = {1 \over \a
(1-\b)}T_{\m\n} ~. \label{matching} \ee

Let us now look at the leading $\mathcal{O}(1/r)$ terms of the
equations of motion. These come from the contributions of terms of
the equations of motion which are multiplied by $L'/L$, or
$\nabla_\m L'/L$. With $\nabla_{\mu}$ we denote covariant
differentiation with respect to the metric $g_{\mu\nu}$. Then,
according to the expansion \reef{Lexp}, these terms contribute as
\be \left.{L' \over L}\right|_{r \to 0}={1 \over r} + {\cal O}(1)
~~~~,~~ \left.{\nabla_\m L' \over L}\right|_{r \to 0}={\nabla_\m \b
\over \b}\cdot{1 \over r} + {\cal O}(1)~. \ee

With the natural assumption that the energy momentum tensor ${\cal
T}_{MN}$ does not blow up close to the distributional singularity
(otherwise the singularity would not be distributional), we obtain
that the ${\cal O}(1/r)$ terms of the  $(rr)$ and the $(r \mu)$
equations yield respectively \bea
 K^{\m\n} \left(G_{\m\n}+W_{\m\n} - {3 \over 2 \a}g_{\m\n}\right)=0~,\label{rr} \\
 {\nabla^\n\b \over \b}\left(G_{\m\n}+ W_{\m\n} - {3 \over 2 \a} g_{\m\n}  \right)  +   \nabla^\n W_{\m\n}=0 ~.\label{rmu}
\eea These two equations act as constraint equations for the two
quantities $K_{\mu\nu}$, $\b$, which have to be determined and
substituted back to the brane Einstein equation \reef{matching}. In
the general case,  there are $11$ independent unknown functions, but
there are $5$ independent constraint equations in \reef{rr},
\reef{rmu}. Therefore, there will in general be functions which
cannot be determined by the local equations around the brane, but
need to be determined by the bulk solution. This, in fact, is
clearly visible in the case of isotropic cosmology on the brane, as
we will see in the next section.

Finally, if we differentiate the brane Einstein equation
\reef{matching} and use the constraint \reef{rmu}, we can derive the
energy conservation equation \be \nabla^\n T_{\m\n} = - {\nabla^\n
\b \over \b (1-\b)} T_{\m\n}~. \label{poulia} \ee From the above, we
see that the energy is not strictly conserved on the brane, but can
radiate in the bulk if the deficit angle changes. We can therefore
anticipate that cosmological evolutions, close to the standard four
dimensional one,  will have $\b$ almost constant. We will come to
this point in a later section.

\subsection{Topological matching conditions: Ring regularisation}

In the previous analysis, we chose an arbitrary smooth
regularisation of the distributional part of the geometry and the
source. In this section, we will analyse a more specific case, where
the codimension-2 source is taken by the limit of a sharp
codimension-1 source with infinitesimal radius. Therefore, we
replace the tip of the cone with a cap glued to the rest of the
compactification with a ring interface (see right part of
Fig.\ref{regularconelambda}). The extrinsic curvature is still
continuous across the ring, but the angular part of the internal
space metric $L$ has a cusp. Inside the cap, since we will finally
take the limit of its thickness to zero, the normal derivatives in
the center of it tend to the values of the corresponding quantities
just inside the ring.

Outside the ring, we assume that the geometry has an expansion as
\be L^+(x,r)\!= \beta(x)r + \sum_{n=2}^\infty {1 \over n!}\b_n(x)
r^n ~, \label{outL} \ee which is the standard conical singularity
structure in the limit of zero ring thickness. On the other hand,
inside the ring, the geometry should tend to two-dimensional flat
space \be L^-(x,r)\!=\!{\cal C}(x,\ep)+r + \sum_{n=2}^\infty {1
\over n!}\bar{\b}_n(x) r^n ~. \label{inL} \ee with the constant
$\displaystyle{{\cal C}(x,\ep)= \ep [\b(x)-1]+\sum_{n=2}^\infty {1
\over n!}(\b_n(x)-\bar{\b}_n(x)) \ep^n}$ to ensure continuity of $L$
across the ring. In the above, $\ep$ is the width of the cap. As we
will see later on, it is important for the consistency of the model
that the higher order coefficients in the expansion of $L$ (at least
$\b_2$) are different inside and outside the ring. The extrinsic
curvature  across the ring is continuous as in the previous
subsection. We will assume that the derivative of the extrinsic
curvature is also continuous, although this will turn up to be a
fact, rather than an assumption, when we study the consistency of
the setup.  The crucial difference with the previous subsection is
that the metric expansion inside the ring \reef{inL} is specified
and $L'$ has a definite jump.

In general, there are two types of distributional singularities that
one can obtain in the $\ep \to 0$ thin limit. One is that  of
codimension-2 which has the structure $\d^{(2)}(r) = {\d(r) \over 2
\pi r}$ and originates from the terms ${\d(r-\ep) \over 2 \pi \ep}$.
There can exist, however, in the same limit a codimension-1 defect
with singularity structure $\d(r)$, which comes simply from
$\d(r-\ep)$. To calculate the distributional pieces of the equations
of motion, we will make use of the following identities where we
denote by ``distr" the distributional parts of the corresponding
quantities \bea {\rm distr}(L''g_{\m\n})&=&(\b-1)[g_{\m\n} +
g'_{\m\n} \ep] \d(r-\ep) \label{dj0}  \,\,\,\, ,\\ \,\,\,\,
\label{d1} {\rm distr}(L''G_{\m\n})&=&(\b-1)[G_{\m\n} + G'_{\m\n}
\ep] \d(r-\ep) \,\,\,\, ,\\ \,\,\,\, \label{d2} {\rm distr} (L''
W_{\m\n}) &=& (\b-1)[ W_{\m\n} + W'_{\m\n} \ep ] \d(r-\ep)~.
\label{d3} \eea Here, we also note the next to leading order terms
in the $\ep$-expansion, since these distributional parts are divided
by $L \sim \ep$ at the ring position. Because of this well defined
ring structure in the internal space, there could be codimension-1
contributions of the equations of motion surviving in the $\ep \to
0$ limit. It is clear to see that the only such possible
contributions come from the $(\m\n)$ equations. To find the
codimension-1 contributions, we should be careful with the
distributional definition of $T_{MN}$ in \reef{co2emt}. This is
because, inside the cap, the energy momentum tensor which becomes
distributional in the $\ep \to 0$ limit, may have an expansion of
the type \be T_{\m\n} = T^{(0)}_{\m\n}+ T^{(1)}_{\m\n}~r + {\cal
O}(r^2) ~. \ee Expanding \reef{co2emt} at $r=\ep$, we obtain \be
T_{MN} = \left[ T^{(0)}_{\m\n} {\d(r-\ep) \over \ep} +
\left(T^{(1)}_{\m\n}  - {1 \over 2} {\b_2 \over \b} T^{(0)}_{\m\n}
\right) \d(r-\ep) \right]{1 \over 2 \pi \b}  \d_M^\m \d_N^\n ~. \ee
In the following we will use $T_{\m\n}$ to denote $T^{(0)}_{\m\n}$.

From the $\d(r-\ep)/\ep$ singular pieces of equations (\ref{6eqs}),
we obtain the matching conditions for the codimension-2 singularity.
Due to the topological origin of the dimensionally extended Lovelock densities,
the  Gauss-Bonnet combination in the bulk theory will give an Einstein brane term
plus extrinsic curvature corrections, at the level of the junction conditions. The latter then
read as before
 \be  G_{\m\n} + W_{\m\n}- {3 \over 2 \a}g_{\m\n} = {1 \over \a (1-\b)}T_{\m\n} ~.
\label{matchingagain}
\ee

From the $\d(r-\ep)$ part of the $(\m\n)$ equations, we obtain the
codimension-1 matching conditions  at $r=\ep$ \be
 \left[G_{\m\n} + W_{\m\n}- {3 \over 2 \a}g_{\m\n} \right]' ={1 \over \a (1-\b)} T^{(1)}_{\m\n}~. \label{new}
\ee As we will comment at the end, we need always have non-zero
$T^{(1)}_{\m\n}$ in order not to overconstrain the system. Let us
note that \reef{new} is a consequence of the ring regularisation.
Such kind of codimension-1 equation did not arise in the the smooth
regularisation in the topological case, since then, only
two-dimensional distributions are well defined.

Let us now look at the $\mathcal{O}(1/\ep)$ terms of the equations
of motion. We will make again the natural assumption that the
non-distributional energy-momentum tensor  $\mathcal{T}_{AB}$ inside
and outside the rings is regular and has no ${\cal O}(1/\ep)$
singularity. In evaluating the  ${\cal O}(1/\ep)$ terms, we should
be careful to do it both infinitesimally outside of the ring and
also infinitesimally inside of it. This is because the function $L'$
has a jump. This function enters in the equations of motion in two
forms, as $L'/L$ and as $\nabla_\m L'/L$. Then, according to the
expansions  \reef{outL} and \reef{inL}, these terms contribute as
\bea
\left.{L' \over L}\right|_+={1 \over \ep} + {\cal O}(1) &,& \left.{\nabla_\m L' \over L}\right|_+={\nabla_\m \b \over \b}\cdot{1 \over \ep} + {\cal O}(1)~, \\
\left.{L' \over L}\right|_-={1 \over \b}\cdot {1 \over \ep}  + {\cal O}(1)&,&\left.{\nabla_\m L' \over L}\right|_-={\nabla_\m \bar{\b}_2 \over \b} + {\cal O}(\ep)~.
\eea

The  $(rr)$ equation has only $L'/L$ pieces, therefore the ${\cal O}(1/\ep)$ terms give us the equation
\be
 K^{\m\n} \left(G_{\m\n}+W_{\m\n} - {3 \over 2 \a}g_{\m\n}\right)=0~.\label{rragain}
 \ee
On the other hand, the $(r \mu)$ equation has both $L'/L$ and
$\nabla_\m L'/L$ pieces. The difference of the contributions of these two pieces is a $\nabla^\n W_{\m\n}$
term, which therefore has to vanish \be \nabla^\n W_{\m\n} =0~.
\label{newcon} \ee The ${\cal O}(1/\ep)$ part of the the $(r \mu)$
equation just inside the ring gives by itself \be {\nabla^\n\b \over
\b}\left(G_{\m\n}+ W_{\m\n} - {3 \over 2 \a} g_{\m\n}  \right)=0 ~.
\label{beta} \ee

Counting again the constraints of \reef{rragain}, \reef{newcon},
\reef{beta}, we find that there are 9 independent equations to
determine the 11 unknown quantities, $K_{\m\n}$ and $\b$. Therefore,
there will again be in general functions which cannot be determined
by the local equations around the brane, but need to be determined
by the bulk solution. However, in the case of isotropic cosmology on
the brane, as we will see in the next section, all the unknown
functions entering in the brane Einstein equation
\reef{matchingagain} are completely determined (modulo integration
constants) and the system therefore closes.

Had we taken $T^{(1)}_{\m\n}=0$ in the codimension-1 matching
conditions \reef{new}, which have 10 independent components, these
matching conditions would act as constraints and therefore in
principle overconstrain  the system. This is indeed the case for the
cosmological ansatz case that we will discuss later.

Finally, let us again differentiate the brane Einstein equation
\reef{matchingagain}.  Using the constraint \reef{newcon}, we see
that the brane energy momentum tensor is always conserved \be
\nabla^\n T_{\m\n} = 0~. \ee Therefore, the ring regularisation,
\ie the fact that we consider not a $C^2$ but a $C^1$ cap
$C_\lambda$ at $r=\epsilon$, freezes brane radiation.

\subsection{General matching conditions}

\begin{figure}[t]
\begin{center}
\epsfig{file=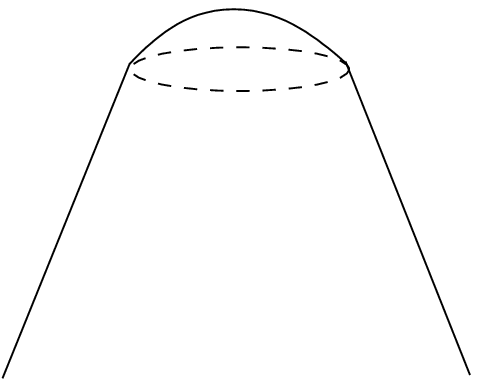,width=5cm,height=4cm}
\begin{picture}(50,50)(0,0)
\SetOffset(15,0)
\Text(-130,110)[c]{$K^-_{\m\n}$} \Text(-145,85)[c]{$K^+_{\m\n}$}
\Text(-45,110)[c]{$L^-$} \Text(-30,85)[c]{$L^+$}
\Text(-90,120)[c]{$r=0$}  \Text(-90,82)[c]{$r=\ep$}
\end{picture}

\end{center}
\caption{The conical singularity regularised by a cap glued to the
bulk at a ring intersurface at $r=\ep$. Both $K_{\mu\nu}$ and $L$
have jumps across at the ring.} \label{regularcone}
\end{figure}

To study the general matching conditions, we will use a similar
regularisation as in the previous subsection, where in addition, the
tangential sector will have in principle distributional jumps. The
regularisation of the codimension-2 singularity is then depicted in
Fig.\ref{regularcone}. In more details, we assume that both the
angular metric function $L$ and the extrinsic curvature $K_{\mu\nu}$
have jumps when going from the exterior to the interior of the ring.
Inside the ring, since we will finally take the limit of its
thickness to zero, the normal derivatives in the center of the cap
tend to the values of the corresponding quantities just inside the
ring.

In this general setup, we would like to investigate the possibility
of having as before a codimension-2 distributional energy-momentum
tensor as in \reef{co2emt}. We consider the case of having a
codimension-1 distributional source as out of the scope of this
paper. The answer to the former question can easily be seen in the
following  to be negative.

The extrinsic curvature inside and outside the ring brane have in
general a non-trivial relation between them, depending on the
details of the internal cap structure. Without loss of generality of
our following analysis, we can make the simplified approximation
that the two extrinsic curvatures are proportional  as \be
K_{\m\n}^-= \e K_{\m\n}^+~, \ee with $\e = {\rm const}$. In the
following, we will simplify the notation denoting with $K_{\m\n}$
the exterior extrinsic curvature $K_{\m\n}^+$ (and similarly for
other quantities constructed by it). Furthermore, we assume that
higher orders in the $r$-expansion of the metric are continuous
across the ring brane. This means in particular that the
non-distributional part of $K'_{\m\n}$ is continuous, in
correspondence with $\b_2$ in the expansion of $L$.

When looking at the distributional part of the equations of motion,
since also $K_{\m\n}$ can now have a jump, we should modify the
distributional parts in \reef{dj0}-\reef{d3} as following \bea {\rm
distr}(L'g_{\m\n})'&=&[(\b-1)g_{\m\n} + (\b - \e)g'_{\m\n} \ep]
\d(r-\ep)  \,\,\,\, ,\\ \,\,\,\, {\rm
distr}(L'G_{\m\n})'&=&[(\b-1)G_{\m\n} + (\b - \e)G'_{\m\n} \ep]
\d(r-\ep)  \,\,\,\, ,\\ \,\,\,\, {\rm distr} (L' W_{\m\n})' &=&
[(\b-\e^2) W_{\m\n} + (\b-1)W'_{\m\n} \ep ] \d(r-\ep)\,\,\,\, ,\\
\,\,\,\, {\rm distr} K'_{\m\n}&=& (1-\e) K_{\m\n}  \d(r-\ep)~. \eea
The difference in the $\ep$-terms in the brackets has to do with the
assumption that $K'_{\m\n}$ is continuous across the ring. All the
derivatives in the above expressions are computed at the exterior of
the ring. From the above we can first write down  the codimension-2
matching conditions from the $\d(r-\ep)/\ep$ terms of the ($\m\n$)
equations of motion \be G_{\m\n} +{\e^2 -\b \over 1-\b} W_{\m\n}- {3
\over 2 \a}g_{\m\n} = {1 \over \a (1-\b)}T_{\m\n}~. \ee For $\e=0$
we recover the matching conditions of \cite{bos}. The codimension-1
$\d(r-\ep)$ part of the ($\m\n$) equations of motion provide us with
the codimension-1 matching conditions which, in agreement with \cite{KS},
give after some simplifications \bea
&&\left[G_{\m\n} + {1-\b \over \e- \b} W_{\m\n}- {3 \over 2 \a}g_{\m\n} \right]' + {3 \b \over 2 \a}~{\e-1 \over \e-\b}(K_{\m\n}-K g_{\m\n})  -\b~{\e^3-1 \over \e-\b}\left(K_\m^\k W_{\k\n}-{1 \over 3}K^{\k \la}W_{\k\la}g_{\m\n}\right) \nn \\
&&+\b~{\e - 1 \over \e - \b} \left[ {\nabla_\k \nabla_\n L \over L}K_\m^\k+{\nabla_\m \nabla_\k L \over L}K_\n^\k -{\Box L \over L}K_{\m\n}-{\nabla_\m \nabla_\n L \over L}K + \left( {\Box L \over L}K -{\nabla_\k \nabla_\la L \over L}K^{\k\la} \right)g_{\m\n} \right] \nn \\
&&+\b~{\e-1 \over \e-\b}\left( R_{\m \k \la \n}K^{\k \la}+G_{\k
\la}K^{\k\la}g_{\m\n}+R_{\m\n}+{1 \over 2}R K_{\m\n}-R_{\k
\n}K_\m^\k -R_{\m \k}K^\k_\n   \right)  ={1 \over \a (\e -\b)}
T^{(1)}_{\m\n} ~. \label{newgeneral} \eea Note that the above
equation for $\e=1$ yields the codimension-1 matching condition
\reef{new} of the previous section.  The codimension-1  $\d(r-\ep)$
part of the $(\th\th)$ equation, on the other hand, will not have an
energy-momentum contribution because we assume that the defect is of
pure codimension-2 nature. Therefore, in agreement with \cite{KS}, we
obtain \be (1-\e)K-{2 \a \over 9}(1-\e^3)K^{\m\n}W_{\m\n}-{2\a \over
3}(1-\e)K^{\m\n}G_{\m\n}=0~.\ \label{ththcon} \ee From the above
constraint equation there are two possibilities. The obvious
solution is $\e=1$, and which gives us the topological matching
conditions. There could be of course some solution for $K_{\m\n}$
satisfying the above equation for $\e \neq 0$. However, as we will
see in the following section for the cosmological ansatz, it turns
out that these codimension-1 parts overconstrain the system.

Regarding the $\mathcal{O}(1/\ep)$ terms of the equations of motion
in this general conditions case, we obtain the constraint
\reef{rragain} from the $(rr)$ equation outside the ring brane \be
 K^{\m\n} \left(G_{\m\n}+W_{\m\n} - {3 \over 2 \a}g_{\m\n}\right)=0~, \label{rrg}
\ee however, inside the ring, because the extrinsic curvature has a
jump, we obtain by substraction of  \reef{rrg} \be \e (\e^2 -
1)K^{\m\n}W_{\m\n} =0~. \label{rrgin} \ee Moreover, from the
$\mathcal{O}(1/\ep)$ terms of the $(r\m)$ equation, we obtain inside
the ring \be {\nabla^\n\b \over \b}\left(G_{\m\n}+ \e^2 W_{\m\n} -
{3 \over 2 \a} g_{\m\n}  \right)=0~, \label{betag} \ee and for the
difference of the corresponing equation outside the brane from the one inside it  \be \nabla^\n
W_{\m\n} =(\e^2-1) {\nabla^\n \b \over \b} W_{\m\n } ~.
\label{newcong} \ee For general $\e$, we see that, apart from the
constraints \reef{rrg}, \reef{betag} and \reef{newcong} that we also
had in the topological ring regularised case, there are 2 more
constraints from \reef{ththcon} and \reef{rrgin}. Thus, since the
total number of constrains is 11, we can in principle determine the
11 unknowns, $K_{\m\n}$ and $\b$. However, it turns out that in
cases of symmetry as the cosmological ansatz that we will study in
the following, the system for general $\e$ becomes overconstrained.

\section{Cosmological equations and consistency}

In this section, we will study the cosmological equations for a
codimension-2 brane. We will first study in which  cases the system
of the brane equations studied in the previous section closes. We
will then move on by checking the consistency of the next order in
the $\ep$-expansion  of the equations of motion. For this purpose we
adopt the following metric ansatz for the  four-dimensional metric
$g_{\m\n}$ describing LFRW cosmology 
\be ds^2 = -n^2(t,r)dt^2 +
a^2(t,r)ds_{\k}^2 + dr^2 + L^2(t,r)d\th^2~, \label{cosmologyansatz}
\ee 
with the spatial
metric for $\k=\pm 1,0$ curvatures \be ds_{\k}^2= {d\r^2 \over 1-\k
\r^2}+ \r^2 (d \f^2+\sin^2 \f ~d\o^2)~. \ee The expansion of $n$ and
$a$ is \bea
n(t,r)&=& 1+ N r + {1 \over 2}N_2 r^2 + {1 \over 6} N_3(x) r^3  +  {\cal O}(r^4)~, \label{nexp}\\
a(t,r)&=& a \left [1 +  A r + {1 \over 2}A_2 r^2 + {1 \over 6}
A_3(x) r^3 + {\cal O}(r^4) \right]~, \label{aexp} \eea where with $a=a(t)$ we
denote the scale factor on the brane.  We will assume that the
energy-momentum tensor on the brane is the one of perfect fluid
$T_{\m}^{\n} = (-\r,P,P,P)$. The energy density and pressure may
contain contributions from the vacuum energy of the brane. We will
not factor out these contributions until the next section.

\subsection{The brane equations of motion}

Let us now write the system of brane equations for the different
cases discussed in the previous section. By the term brane
equations, we mean the matching conditions supplemented with the
constraints from the ${\cal O}(1/ \ep)$ parts of the  $(rr)$ and
$(r\m)$ equations, and the codimension-1 constraints when
appropriate.

For the case of topological boundary conditions with a smooth regularisation, we obtain that the equations
\reef{matching}, \reef{rr}, \reef{rmu} give the following system
\bea
\frac{\rho}{3\alpha (1-\beta)}&=& H^2+\frac{\kappa}{a^2}+\frac{1}{2\alpha}-A^2 ~, \label{00cosmo} \\
-\frac{\rho+3P}{6\alpha (1-\beta)}&=& \frac{\ddot{a}}{a} +\frac{1}{2\alpha}-f A^2 ~, \label{ijcosmo}\\
f\equiv {N \over A}&=&\frac{3P}{\rho} ~, \label{rrcosmo} \\
2A \dot{A}+2H(1-f)A^2&=&\frac{\dot{\beta}}{\beta}\frac{\rho}{3\alpha
(1-\beta)} ~. \label{rmucosmo} \eea In order to derive the equations
\reef{rrcosmo} and \reef{rmucosmo}, we have used the first one
\reef{00cosmo}. The above system of equations has evidently one free
function, which we can take to be $\b(t)$. This function is expected
to be fixed by the  asymptotic dynamics of the bulk and cannot be
determined by the local equations of motion.

For the case of topological boundary conditions with ring-type
regularisation, we  find that \reef{rmucosmo} is split in two parts.
The equations \reef{newcon} and \reef{beta} now give \bea
2A \dot{A}+2H(1-f)A^2&=&0 ~, \label{rmucosmor} \\
\dot{\b}~\r&=&0 ~, \label{Wcosmo} \eea From the last one, we will in
the following choose the solution $\dot{\b}=0$, i.e. that the
deficit angle is constant, since the other possible solution demands
that the brane energy density vanishes ($\r=0$). In this case, the
system of \reef{00cosmo}, \reef{ijcosmo}, \reef{rrcosmo},
\reef{rmucosmor} and \reef{Wcosmo} is closed, the cosmology on the
brane is uniquely determined, and is not depending on the bulk
dynamics but only on the local structure of the equations of motion.

The general boundary conditions  on the other hand, have
corresponding equations with the ones of the previous case, plus two
more which have to do with the constraint \reef{rrgin} and the
codimension-1  constraint \reef{ththcon}. These read \bea
\label{mpekris}
\frac{\rho}{3\alpha (\eta^2-\beta)} & = & \frac{1-\beta}{\eta^2-\beta}(H^2+\frac{\kappa}{a^2}+\frac{1}{2\alpha})-A^2  \\
-\frac{\rho+3P}{6\alpha (\eta^2-\beta)} & = & \frac{1-\beta}{\eta^2-\beta}(\frac{\ddot{a}}{a} +\frac{1}{2\alpha})-f A^2 \\
f\equiv {N \over A} & = & {3 P -3 \a A^2(\e^2-1) \over \r + 6 \a A^2 (\e^2-1) } \\
2A \dot{A}+2H(1-f)A^2 &=& \frac{\dot{\beta}}{\beta} (\e^2-1) A^2 \\
\dot{\b} \left(\r + 3 \a \b (1-\b) (\e^2-1) A^2 \right) &=& 0 \\
\e (\e^2-1)(1+2f) &=& 0 \label{rrgincos} \\
(1-\e)\left[f+3-{2\a \over 3}(1+\e+\e^2)A^2(1+2f)+ 2\a H^2 (1+f) -4
\a {\ddot{a} \over a} \right]&=&0 .\label{ththconcos} \eea
Obviously, for the topological case we arrive at the generic
conclusions we explained in the previous section. Let us concentrate
on the general case. From \reef{rrgincos} we see that can either
have $\e=0$ or $f=-1/2$ with general $\e$. The choice $\e=0$ is
inconsistent for general matter on the brane because of the
codimension-1 constraint \reef{ththconcos}.  On the other hand the
choice $f=-1/2$ fixes the equation of state to $w=-1/6$ (for tensionless branes - for branes with tension the equation of state will be fixed to another constant value), therefore
it is again not so interesting.

From the above we deduce that the only consistent boundary
conditions with a purely codimension 2 defect are the topological
ones. This theorem lies on two assumptions: first that the cosmological ansatz \reef{cosmologyansatz} does 
not contain any $\verb"g"_{tr}$ or $\verb"g"_{r\th}$ components (see \cite{wett} for criticism of the \reef{cosmologyansatz} ansatz) and second that the metric is analytic at $r=0$ so that it can be expanded in powers or $r$ as in \reef{outL}, \reef{inL}, \reef{nexp}, \reef{aexp}.  Let us also stress that, if we had had a true codimension-1 defect
with non-zero $T_{\th\th}$, there would be no consistency issue {\it
for any} $\e$, since then the constraint \reef{ththconcos} would
have a matter part in the right hand side.

\subsection{Consistency of the next order equations} \label{consistencysection}

Up to this point, we have looked at the matching conditions
supplemented by some equations of order $1/\ep$.  These included
apart from the brane metric, the deficit angle $\b$ and the
extrinsic curvature $K_{\m\n}$. In the cosmological case the
corresponding functions were the following: $a$, $\b$, $A$ and $N$.
It is easy to see that in general the various terms in the
$\ep$-expansion of the equations of motion break into groups
according to the unknown functions that they involve. The resulting
grouping can be easily seen from Fig.\ref{matrix}.

The next group of equations (which we call second order) is the one
with the  ${\cal O}(1/\ep)$ terms of the $(\m\n)$ equations and the
${\cal O}(1)$ equations of the $(\th\th)$, $(rr)$ and $(rt)$. These
form a set  of  five equations which involve the next order
variables $A_2$, $N_2$ and $\b_2$. We will study them in the
cosmological case to see if they are consistent. The strategy is to
solve for them from three of these equations and then check that the
two extra equations do not overconstrain the system. This check is
done in view of \cite{thereof}, where it was claimed that the system
is inconsistent (overconstrained). Here, we find that a careful
inclusion of the $(L'/L) K_{\m\n}$ terms omitted in \cite{thereof}
gives consistency. Without loss of generality, we will make the
simplifying assumption for this proof that the bulk is empty, \ie
${\cal T}_{MN}=0$, although some matter could be necessary in the
interior of the defect for the ring regularised case. We will comment at the end of the section
about what happens if we drop this assumption.

\begin{figure}[t]
\begin{center}
\begin{tabular}{|c|c|c|c|}  \hline
  ~~~~~~~~~~  & ~~~~JC~~~~ &  ~~${\cal O}(1/\ep)$~~ & ~~~${\cal O}(1)$~~~ \\  \hline
(rt) &  & $\bigstar$ & $\blacksquare$ \\  \hline
(rr) &  & $\bigstar$ & $\blacksquare$ \\  \hline
($\th \th$) & &  &  $\blacksquare$ \\ \hline
(tt) & $\bigstar$ & $\blacksquare$ & $\blacktriangle$ \\  \hline
(ij) & $\bigstar$ & $\blacksquare$ & $\blacktriangle$  \\  \hline
\end{tabular}
\hspace{5mm}
\begin{tabular}{|c|c|}  \hline
  ~~~~~~~~~~  & ~~~~Functions~~~~  \\  \hline
$\bigstar$  &  $a$, $N$, $A$, $\b$ \\  \hline
$\blacksquare$ &  $N_2$, $A_2$, $\b_2$, $\bar{\b}_2$ $+$ the above \\  \hline
$\blacktriangle$ & $N_3$, $A_3$, $\b_3$, $\bar{\b}_3$ $+$ the above \\ \hline
\end{tabular}
\end{center}
\caption{The grouping of equations according to the various functions that are to be determined. On the left table, the first column has the codimension-2 junction conditions, while the other columns have the various orders in $r$ of the Einstein equations. The ${\cal O}(1/\ep)$ component of the  $(\th \th)$  Einstein equation vanishes identically.} \label{matrix}
\end{figure}

Let us look first for the topological case without a specified regularisation.  We can first solve for $A_2$, $N_2$ and $\b_2$ from the ${\cal O}(1/\ep)$ parts of $(tt)$ and $(ij)$ equations and from the   ${\cal O}(1)$ part of $(\th\th)$ equation. As it can be seen in Appendix \ref{2ndorder}, these equations of motion form a $3 \times 3$ system of linear algebraic in the second order variables equations. Using the brane equations \reef{00cosmo}-\reef{rmucosmo}, we can easily solve for $A_2$, $N_2$ and $\b_2$ as
\bea
A_2&=&\frac{15(1-\b)^2+2(1-\b)(\r+9P)+4\r P - {2 \r^2 \over A^2}{\dot{\b}^2 \over \b^2}}{4 \a (1-\b)(\r+9P)} \\
N_2&=&\frac{-\r(\r+3P)^2-{45 \over 2}(1-\b)^2(\r+6P)+3(1-\b)\r(\r+9P)-6\r^2 P + { \r^3 \over A^2}{\dot{\b}^2 \over \b^2}}{6 \a (1-\b)\r(\r+9P)}\\
\b_2&=&\b A\frac{45(1-\b)^2-2\r(\r+3P) - {6 \r^2 \over
A^2}{\dot{\b}^2 \over \b^2}}{2 \a \r(\r+9P)}. \eea \noindent The
next step is to make sure that the ${\cal O}(1)$ parts of $(rr)$ and
$(rt)$ equations do not overconstrain the system. These equations
are written down in Appendix \ref{2ndorder}. It is a tedious, but
straightforward exercise to see that in fact they are automatically
satisfied and therefore the system is consistent.

The next check we have to do concerns the ring regularisation in the
topological matching conditions. Here we need to compute the
equations also in the interior in order to see if the system is
consistent. In this case, a simplification which will help us is
that the equations for the first order variables imply that
$\dot{\b}=0$. This simplifies considerably  the equations. In the
interior, having assumed that $A_2$ and $N_2$ are the same
(continuity of $K'_{\m\n}$), we have the extra variable
$\bar{\b}_2$. The algebraic system in the interior of the   ${\cal
O}(1/\ep)$ parts of $(tt)$ and $(ij)$ equations and from the ${\cal
O}(1)$ part of $(\th\th)$ equation, is the same as for the exterior
with the only substitution \be \bar{\b}_2={ \b_2 \over \b }\,\,.
\label{barb2} \ee \noindent The next step for the consistency of the
system in the interior is to see the ${\cal O}(1)$ parts of $(rr)$
and $(rt)$ equations. The ${\cal O}(1)$ part of the $(rt)$ equation
is indeed then identically satisfied. However, the ${\cal O}(1)$
part of $(rr)$ equation is not satisfied without the inclusion of a
${\cal T}^{(in)}_{rr}$ matter in the interior. In more detail, the
matter we have to include in the interior of the ring in order not
to overconstrain the system is given by the latter equation  to be
\be {\cal T}^{(in)}_{rr}=
\frac{-45(1-\b)^2+6(1-\b)(\r-3P)-2\r(\r+3P)}{24\pi \a \b (1-\b)}
\equiv X \,. \label{trr} \ee Had we not taken a ${\cal
T}^{(in)}_{rr}$, then the equations would only be satisfied for a
specific relation of $\r$ and $P$ of the brane matter. This would
lead to overconstraining the system, and in the spirit of
\cite{thereof} to inconsistency. However, it is natural to assume
that the thickening out of the thin codimension-2 defect gives rise
to a non-zero energy momentum in the interior of it.

As far as the third order equations of the table in Fig.\ref{matrix}
are concerned, there is no question about consistency. This is
because there are three new variables $A_3$, $N_3$ and $\b_3$ with
two equations (in the smooth regularisation case), the ${\cal O}(1)$ parts of the $(tt)$ and the $(ij)$
equations. In the ring regularisation case, there is one more variable $\bar{\b}_3$, but two more
equations, as the number of equations is doubled (evaluated inside and outside the ring), so again
every function can be determined. Therefore, we will not study them in more details.

Finally, let us comment about the differences in the above approach
if we had allowed for general matter ${\cal T}_{MN}$ in the bulk. It
is obvious that in this case the above solutions for the second
order variables are altered. Then, for the smooth regularisation, we
would obtain from the ${\cal O}(1)$ parts of the $(rr)$ and $(rt)$
equations that \be {\cal T}_{r\th}=0 \, \, \, , \, \, \, {\cal
T}_r^r = {\cal T}_\th^\th ~. \ee The same would hold for the
exterior to the ring matter in the ring regularisation case. For the
interior, the equations of motion are satisfied if \be {\cal
T}^{(in)}_{r\th}=0 \, \, \, , \, \, \, {\cal T}_\th^{(in)~\th} =
{\cal T}_\th^\th  \, \, \, , \, \, \,  {\cal T}_r^{(in)~r} = X + {1
\over \b}{\cal T}_\th^\th ~, \ee where $X$ is the interior $rr$
component of the energy-momentum tensor in the absence of other
matter given in \reef{trr}.

\subsection{Vanishing extrinsic curvature limit} \label{vanishing}

In the previous section, one could wonder about what happens in the
limit of vanishing extrinsic curvature ($K_{\m\n}$), since several
expressions are divided by $A$ or $N$. In this subsection, we will
look back to the original equations without dividing them by
components of the extrinsic curvature. First the brane equations
read for both types of regularisation that we considered \bea
\label{bustop}
\frac{\rho}{3\alpha (1-\beta)}&=& H^2+\frac{\kappa}{a^2}+\frac{1}{2\alpha} ~,  \\
\label{bustop2}
-\frac{\rho+3P}{6\alpha (1-\beta)}&=& \frac{\ddot{a}}{a} +\frac{1}{2\alpha} ~, \\
\dot{\beta}~\r&=&0 ~.
\eea

From the above we have two distinct choices. First that $\r=0$.
Differentiating the Friedmann equation and using the acceleration
equation, we find then that also  $P=0$. The solution then for the
scale factor is de-Sitter space in all possible foliations ($\k=\pm
1,0$). The second order constraints for the ${\cal O}(1)$ parts of
the $(00)$, $(ij)$ equations and the ${\cal O}(1)$ part of the
$(rt)$ equation are trivially satisfied. On the other hand, the
${\cal O}(1)$ parts of the $(\th\th)$ and the $(rr)$ equations are
only satisfied with the inclusion of bulk matter. In both types of
regularisations (smooth and ring)  and for both inside and outside the ring in the ring
regularisation case, the later equations give \be {\cal
T}_r^r={\cal T}_r^r= {15 \over 8 \pi \a} ~, \ee which corresponds
(when also looking at the ${\cal O}(1)$ parts of the $(\m\n)$
equations) to cosmological constant $\La_6=-5/2\a$. This value of
the bulk cosmological constant  in fact realises the Born-Infeld
limit \cite{zanelli} as noted in \cite{papapapa}.

The second distinct case is when $\dot{\b}=0$. Notice that then, the
equations (\ref{bustop}), (\ref{bustop2}) are the 4-dimensional
standard LFRW equations{\footnote{Parts of these observations can be
found in the PhD thesis of Paul Bostock \cite{bus2}.}}. Then, the
second order constraints for the ${\cal O}(1)$ parts of the $(00)$,
$(ij)$ equations and the ${\cal O}(1)$ part of the $(rt)$ equation
are trivially satisfied for $\b_2=\bar{\b}_2=0$. On the other hand,
the ${\cal O}(1)$ parts of the $(\th\th)$ and the $(rr)$ equations
give constraints. In the smooth regularisation, they coincide and
give a relation between $A_2$ and $N_2$. In the ring regularisation,
the same holds for the outside problem. Inside the ring,  the ${\cal
O}(1)$ part of the $(\th\th)$ equation gives the same relation
between $A_2$ and $N_2$. On the other hand, the ${\cal O}(1)$ part
of the $(rr)$ equation needs some extra ${\cal T}_{rr}$.

It is instructive to see the limit of this special case for
Minkowski brane vacua, where $H=0$ and $\k=0$, with cosmological
constant $\La_6$ in the bulk. Then, the brane has tension $T_4= \r =
-P =3 (1-\b)/2$. In the smooth regularisation case, the relation
between  $A_2$ and $N_2$ is rather simple \be N_2= -3A_2 - \La_6 ~.
\label{bhtest} \ee In the ring regularisation, the  ${\cal O}(1)$
part of the $(rr)$ equation dictates that \be {\cal T}^{(in)}_{rr}=
-{3 \over 4\pi}\left[ \La_6 +  \La_6 {1-\b \over \b} \right]~, \ee
where the first addendum is the bulk cosmological constant
contribution and the second is an extra contribution necessary for
the regularisation to work. We can compare then, this special case
with the known exact solution of \cite{papapapa} of the double Wick
rotated black hole. As we see in Appendix \ref{bh}, the above
reproduce the correct relation between the bulk cosmological constant and
the coefficients of $K'_{\m\n}$. This example is important, because
it violates the assumption of the form of the  expansion of
$K_{\m\n}$ that was considered in \cite{bos} (namely $K'_{\m\n}=0$).
The same assumption was subsequently used in \cite{pap2} and led to
erroneous constraints for the brane matter.

\section{Codimension-2 cosmological evolution}

Let us now revisit the brane equations and try to understand what
kind of cosmological evolution can be obtained on the codimension-2
brane. Let us stress here that, although we proved that the equations of motion locally around the purely codimension-2 defect are consistent for the topological boundary conditions, one has to make sure that there are no singularities introduced when integrating the equations of motion away from the defect. Here, we will not carry out this formidable task of the bulk integration, but we will content 
ourselves with deriving the possible four dimensional  cosmologies from just the local equations on the brane. As discussed in the previous sections, the brane equations of
motion for the smoothly regularised topological case are \bea
\frac{\rho}{3\alpha (1-\beta)}&=& H^2+\frac{\kappa}{a^2}+\frac{1}{2\alpha}-A^2 ~, \label{mpekris00} \\
-\frac{\rho+3P}{6\alpha (1-\beta)}&=& \frac{\ddot{a}}{a} +\frac{1}{2\alpha}-f A^2 ~, \label{mpekrisij}\\
f\equiv {N \over A}&=&\frac{3P}{\rho} ~, \label{founta} \\
2A \dot{A}+2H(1-f)A^2&=&\frac{\dot{\beta}}{\beta}\frac{\rho}{3\alpha
(1-\beta)} ~. \label{kkees} \eea Here, we recognize the first two
equations as a modified version of the four-dimensional Friedmann
and acceleration equations. Note that $\beta$ is generically a
function of time which can make the effective four dimensional
Planck scale time varying. A second important point is the
appearance of an effective cosmological constant in the face of
$\alpha$, which is of geometric origin. A last difference from the
standard four dimensional equations is  the presence of the
extrinsic curvature correction parametrised by $A$.

On the other hand, the continuity equation \reef{poulia}, which is
not independent from the above, gives \bea \dot{\rho}+3H(\rho+P) +
\frac{\r~\dot{\b}}{\b(1-\beta)}=0.\label{kke} \eea Note that, in
general, there is not energy conservation for the brane matter since
a varying $\beta$ means inevitable energy exchange between the bulk
and brane. This is in contrast with the  $\eta=0$ case where there
is energy conservation \cite{kofi1}. The energy conservation
equation \reef{kke} can be rewritten as \be
\frac{d}{dt}\left(\frac{\rho
\beta}{1-\beta}\right)+3H\frac{(\rho+P)\beta}{1-\beta}=0,\label{kkemulu}
\ee which looks like the standard energy conservation equation for a
redefined energy density and pressure as $(\r,P) \to
(\r,P)\b/(1-\b)$. In fact, the five (four independent) coupled
equations of motion, \reef{mpekris00}, \reef{mpekrisij},
\reef{founta}, \reef{kkees}, \reef{kke}, which are the full
information available to us at $r=0$, are consistent with each other
and  constitute a non-closed system. In other words, one needs extra
information coming from the bulk geometry in order to fix one of the
functions, e.g. $\beta$, and then to solve fully the system. The
solution in the bulk is no longer unique as in the case of
codimension-1 brane cosmology and one has a family of bulk solutions
parametrised by the angular deficit function $\beta$. Not all these
bulk solutions will be acceptable. Certain of them will inevitably
carry singularities away from the brane for example, and not only in
the position of the brane at $r=0$.

From the matching conditions \reef{matching}, we can read the
effective four-dimensional gravitational constant  as \be
\label{blank} \k_{4}^2=\left({3 \k_6^2 \over 4 \pi}\right) {1 \over
\alpha (1-\beta)}~, \ee where we have momentarily reintroduced the
six-dimensional gravitational constant according to (\ref{6action}).
As noted before, an energy exchange between bulk and brane means a
time-varying gravitational constant through the variation of $\b$.
This variation is constrained during the early cosmology  by the
primordial abundances at the nucleosynthesis epoch, with a limit of
this variation approximately  $\frac{|\dot{G}|}{G H}\lesssim 0.2$
for $G=8\pi \k_4^2$ (see \cite{dG} for details of this limit). This
is constraining the  variation of $\b$ as \be \left|{\dot{\b} \over
(1- \b)H}\right| \lesssim 0.2~, \ee which is not a rather strong
constraint. Further constraints come from the fact that the theory
with varying $\b$ is similar to a scalar-tensor theory and therefore
there will be strong constraints from solar system observations. Not
knowing the full family of solutions in the bulk, we choose to
consider the case where $\b$ is {\it approximately constant}. This,
in fact, will be exactly the case if we restrict our analysis to the
ring  topological regularisation, where there is no energy exchange
between the defect and the bulk. This case is bound to give us, at
least seemingly, acceptable four-dimensional cosmology, and has the
merit that the system of equations is closed and does not depend on
undetermined (by the matching conditions) functions.

When $\b$ is constant, we can solve for the extrinsic curvature $A$
for a given, constant equation of state for the brane matter. For that
purpose, we split the energy density and pressure to a part which
has the form of brane vacuum energy $\la$ and part which describes
brane matter \be \r= \la+ \r_m \ \ \ \ , \ \ \ \ P=-\la +P_m~, \ee
with the brane matter equation of state defined as $w=P_m/\r_m$.
Then, for $w \neq -1$, we can integrate \reef{kkees} as \be A^2 =
{C^2 \over (\la+\r_m)^2}~\r_m^{8 \over 3(1+w)}~, \label{Asolution}
\ee with $C^2$ a positive integration constant. With this
expression, the Friedmann  and  acceleration equations
\reef{mpekris00}, \reef{mpekrisij} become \bea
H^2+\frac{\kappa}{a^{2}}&=&
{\k_4^2 \over 3} \r_m +\left({\k_4^2 \over 3}\la  - \frac{1}{2\alpha} \right) +
{C^2 \over (\la+\r_m)^2}~\r_m^{8 \over 3(1+w)}~,  \label{modH} \\
\frac{\ddot{a}}{a} &=& - {\k_4^2 \over 6}(1+3w)\r_m +\left({\k_4^2
\over 3}\la  - \frac{1}{2\alpha} \right) + 3 {C^2 \over
(\la+\r_m)^3}(w\r_m - \la)\r_m^{8 \over 3(1+w)} \label{modqacc} ~,
\ee which show a non-trivial correction to the cosmological
equations as a function of $w$. The magnitude of this correction
depends on the integration constant  $C^2$. In the following, we
will try to see whether it is possible for large enough $C^2$,
consistent with observations, to have some interesting modification
to cosmology. From the integration of the continuity
equation \reef{kkemulu} for $\b=$const. we obtain \be \r_m = {\r_0 (1-\b) \over \b
a^{3(1+w)}} ~,  \label{rhoa} \ee which holds also for the special $w=-1$ case.

For $C^2 \neq 0$, our six-dimensional bulk geometry has a genuine
curvature singularity at $r=0$ (apart from the distributional one).
In fact, this is to be expected from purely geometrical
considerations (see \cite{cosmicpbranes}). Higher codimension
defects, when considered in their zero width limit, will develop
curvature singularities. These are expected to be smoothable once we
take finite width corrections into account. Therefore, we expect
finite width effects to be important at the UV sector and our
distributional approximation to break down even though this will not
show up necessarily in the field equations themselves!

Note that equations \reef{modH}, \reef{modqacc} are valid not only
for $\beta$ exactly constant, but also for a $\beta(t)$ which is slightly
oscillating around a constant value. To see this, we note from
\reef{kkemulu} that (as far as $\rho_{m}+P_{m}>0$) the quantity
$u=\rho\beta/(1-\beta)$ has opposite monotonicity than $a$, and
therefore, $u$ can be used as a good time parameter. Note that for this case
the solution \reef{rhoa} is not valid in general. From equations
\reef{mpekris00}-\reef{kkees}, we can obtain a differential equation
for $d(H^{2})/du$ containing also the function $\beta$. Using the
small oscillatory behaviour of $\beta$, we get an autonomous
equation for $d(H^{2})/du$, whose integration gives equations
\reef{modH}, \reef{modqacc}.

Let us now analyse some particular cases of interest.

\subsection{Self-accelerating branes}

The first interesting case is the one in which the tension of the
brane is vanishing $\la=0$. In this case, we see from \reef{modH}
that there is an {\it asymptotic accelerating phase} for  $\a<0$.
This acceleration is purely due to the geometric Gauss-Bonnet term
and therefore is a case of  self-acceleration. This corresponds to
generalisation of the self-accelerating solutions of \cite{papapapa}
with the addition of matter. Before proceeding, let us note that in order  
that these self-accelerating solutions are viable, one has to check that they are free 
from ghosts that are typical in five dimensional self-accelerating braneworlds  \cite{ghost}. The
correction to the Friedmann equation is given by \be A^2=
C^2~\r_m^{2(1-3w) \over 3(1+w)} = {\tilde{\C}^2 \over a^{2(1-3w)}
}~. \label{Arho} \ee Note that the second equality holds also for
$w=-1$. In the table in Fig.\ref{eos} we list the form of
corrections that we obtain for different constant equations of
state. It is worth noting four interesting limits of the
cosmological evolution in this case.

\begin{figure}[t]
\begin{center}
\begin{tabular}{|c|c|c|}  \hline
   ~~~~~~$w$~~~~~~ &  ~~~~~~$\r_m$~~~~~~ & ~~~~~~$A^2$~~~~~~ \\  \hline
 $-1$ & const. & $a^{-8}$ \\  \hline
 $-2/3$ & $a^{-1}$ & $a^{-6}$ \\  \hline
 $-1/3$ & $a^{-2}$ & $a^{-4}$  \\  \hline
 $0$ & $a^{-3}$ & $a^{-2}$ \\  \hline
 $1/3$ & $a^{-4}$ &  const. \\ \hline
 $1$ & $a^{-6}$ &  $a^4$ \\ \hline
\end{tabular}
\end{center}
\caption{The functional dependence on the scale factor of the energy density and the extrinsic curvature correction to the Friedmann equation for different  equations of state of the brane matter in the self-accelerating case. From top to bottom we have the equations of state for cosmological constant, domain walls,  cosmic strings, dust, radiation and stiff matter.} \label{eos}
\end{figure}

First is the one where the brane is dominated by cosmological
constant. This may go against the initial assumption that the
tension of the brane vanishes, because of the ambiguity to separate
the vacuum part of the energy density from the matter energy
density. If a matter component behaving as inflaton or dark energy
dominates at some period of the history of the Universe, it will
have the behaviour that we note here. So, for this case the
extrinsic curvature plays the role of a dark radiation squared term
$a^{-8}$ that will dominate the cosmology at early times.

A second interesting limit is when the extrinsic curvature on the
brane vanishes $A=0$. This limit is of particular interest since we
have a regular six-dimensional geometry at the location of the brane
at $r=0$. Then, the  codimension-2 cosmology is exactly the
four-dimensional LFRW plus an effective geometric cosmological
constant $\frac{1}{2\alpha}$ which is positive for negative
$\alpha$. This is the generalised version of the solutions of
\cite{papapapa} which were obtained  in the case of pure tension. We
see clearly that even in the absence of matter $\rho=0$ and
$\kappa=0$ we have a non-zero $H^2$, hence these solutions are
self-accelerating.

\begin{figure}[t]
\begin{center}
\epsfig{file=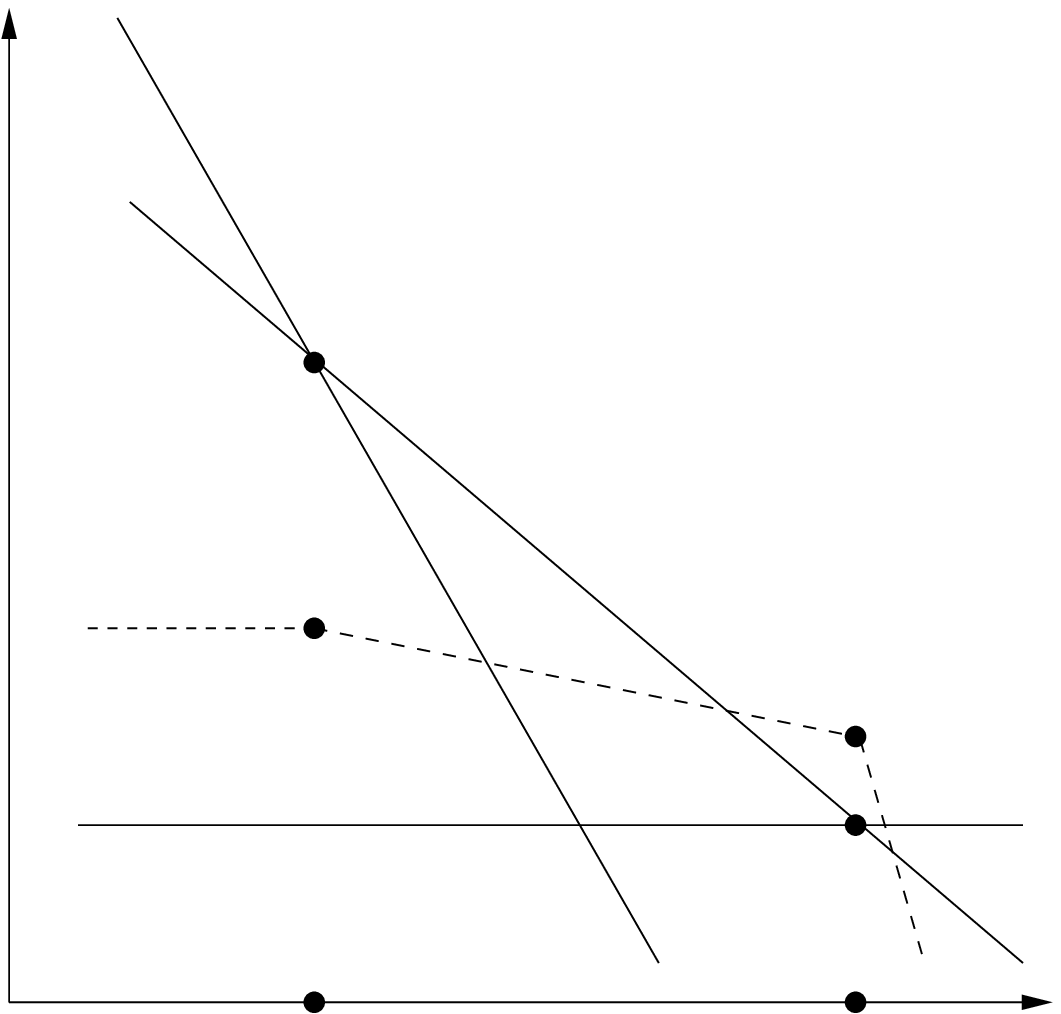,width=7.5cm,height=6cm}
\begin{picture}(50,50)(0,0)
\SetOffset(15,0)
\Text(-250,160)[c]{$\ln \r_m$} \Text(-10,-10)[c]{$\ln a$}
\Text(-190,40)[c]{$1/(2|\a|)$} \Text(-190,110)[c]{dust}  \Text(-170,140)[c]{rad.}
\Text(-170,-10)[c]{$\ln a_{eq}$}  \Text(-55,-10)[c]{$\ln a_{acc}$}
\Text(-190,72)[c]{const.} \Text(-110,48)[c]{$a^{-2}$} \Text(-55,13)[c]{$a^{-8}$}
\end{picture}
\end{center}
\caption{The log-log evolution of the  energy densities as a function of the scale factor. With solid lines, there are the contributions of the standard matter and the geometrically induced $1/(2|\a|)$ cosmological constant. With dashed lines are the evolutions of the extra component ($\propto A^2$) in the Friedmann equation due to the extrinsic curvature of the brane. The dependence of this extra component on the scale factor is noted. This dependence changes whenever a new matter component becomes dominant, \ie at matter domination at $a_{eq}$ and at the cosmological constant domination at $a_{acc}$.} \label{density}
\end{figure}

A third interesting limit is when the matter equation of state is
that of radiation $w=1/3$. Then, the extrinsic curvature correction
is of the form of constant vacuum energy. One would be tempted to
use this vacuum energy to drive early Universe inflation. This seems
to be possible, but around  matter-radiation equality  the Universe
will be dominated by this new vacuum energy, making the
phenomenology of the model problematic.

A fourth interesting case is when the matter equation of state is
that of dust $w=0$. Then the extrinsic curvature correction is of
the form of curvature ($a^{-2}$). In this case, we can see a
possibility of having an observable signature of the codimension-2
cosmology, if the constant $C^2$ is chosen so that there is a brief
period of curvature domination around the vacuum energy domination
period. This scenario is depicted in Fig. \ref{density}. In order to
obtain the standard epochs, \ie matter domination and then
cosmological constant domination,  we need to have that
$\r_m(a_{eq})> 1/(2|\a|)$ and $(\k_4^2/3)\r_m(a_{eq})> A^2(a_{eq})$,
where  $a_{eq}$ is the scale factor at radiation-matter equality and
$a_{acc}$ the one at the moment of cosmological constant domination.
In order that the brief period of curvature domination is
observable, we need $A^2(a_{acc})> 1/(2|\a|)$. Therefore combining
these inequalities we see that we need \be {1 \over 2|\a|} <
A^2(a_{acc}) < {\k_4^2 \over 3}\r_m(a_{eq})~. \ee Then, a
straightforward manipulation with the help of \reef{rhoa},
\reef{Arho} provides us with the following constraint on $C^2$ \be
\left(\r_0 (1-\b) \over \b \right)^{-2/3} {a_{acc}^2 \over 2 |\a|} <
C^2 < {\k_4^2 \over 3}\left(\r_0 (1-\b) \over \b \right)^{1/3}\left(
a_{acc}^2 \over a_{eq}^3 \right)~. \ee It is not yet clear if such a
brief period of curvature domination is phenomenologically viable
and it requires certainly further study to see the potential
observational signatures of such a case.

\subsection{Self-tuning branes}

A second interesting case is when self-tuning can be realised. For
that possibility, the topological quantity $\b$ is used to
accurately cancel the vacuum energy contribution of the brane to the
effective cosmological constant. In more details, as discussed in \cite{papapapa}, by tuning $\b$ one
can make the two vacuum energy contributions, the geometrical one
$1/2|\a|$ and the brane tension one $\la$,  cancel 
\be {\k_4^2 \over
3}\la  - \frac{1}{2\alpha}=0~. \label{betacon} \ee 
This  limit is
orthogonal to the self-accelerating case that we studied before. Such kind of self-tuning
solutions, found in  \cite{papapapa}, have of course to be checked for their stability.  The
effective Friedmann equation for the present case  is given by the
expression \be H^2+\frac{\kappa}{a^{2}} = {\k_4^2 \over 3} \r_m  +
{C^2 \over ({3 \over 2\a \k_4^2}+\r_m)^2}~\r_m^{8 \over 3(1+w)}~,
\ee which includes a non-trivial correction beyond the standard
linear to energy density term. Let us note here that the actual
self-tuning is not visible from this Friedmann equation. This is
because the present is valid for $\b={\rm const.}$ Instead, one
should consider the time-varying deficit angle case in order to see
how \reef{betacon} can be dynamically achieved at the early Universe
evolution. Since before nucleosynthesis there are no constraints on
the variation of Newton's constant, one could have an acceptable
cosmology with varying $\b$. However, then, the brane equations do
not close and the mechanism of self-tuning should be  dictated by
bulk boundary conditions. The study of such a case is beyond the
scope of the present paper.

\section{Conclusions}

In this paper we have demonstrated that a distributional treatment
of codimension-2 branes is consistent and possible if one considers
the full Lovelock gravity in six dimensions. Higher dimensional GR
cannot describe infinitesimally thin codimension-2
defects{\footnote{There is a slight caveat in this argument as it is
unknown mathematically if breaking of axial symmetry may remedy this
problem.}} for it does not present the differential complexity to
allow such distributional terms \cite{cz}. Furthermore, we have
shown that it is the higher order gravity terms in the bulk action
that guarantee the presence of ordinary GR gravity on the brane.
This, up to now, was hinted by the general form of the junction
conditions \cite{bos}, \cite{cz}, but never demonstrated as a
consistent solution of the full bulk and brane field equations. A
nice example where the importance of the full bulk field equations
becomes essential is the ``apparent"  choice of junction conditions
\cite{cz}. In principle, at the level of distributions, differing
mathematical regularity can give differing matching conditions, both
a priori completely consistent. It is only upon considering the bulk
field equations (in the spirit of \cite{KS}) that we see that the
general junction conditions introduced in \cite{bos} inevitably need
codimension-1 matter sources. We have therefore established a
genuine physical difference between the topological and general
junction conditions. The former are pure codimension-2 junction
conditions, whereas the latter are mixed codimension-1 and
codimension-2 junction conditions. It would be interesting to see the relation of the general matching
conditions with work on intersecting branes and cascading cosmologies \cite{casc}.

We have studied the cosmology of conical branes and have found
several interesting features. Firstly, quite generically the higher
order terms give ordinary LFRW equations, as though one included  an
induced gravity term on the brane \cite{DGP}. This is quite different
from codimension-1 cosmology \cite{bow}, where the ordinary behaviour
is recovered only at late times. Corrections to this
standard evolution are threefold: there is a geometric acceleration
scale related to $1/\alpha$, which restates the problem of a minute
cosmological constant as a gravitational hierarchy between the bulk Gauss-Bonnet
and Einstein-Hilbert term. Secondly, extrinsic curvature corrections
are apparent and they are dependent of the equation of state of the
perfect-fluid matter. In other words, as the brane evolves in the
bulk, a dust equation of state can lead to an extra component
behaving as curvature, or a radiation equation of state to a
cosmological constant term, and so on. To put it in a nutshell, each
matter fluid introduces two differing fluid components in the
modified LFRW equations. Last, but not least, as the brane evolves it
can generically radiate in the bulk. Again, this is unlike the vacuum
bulk codimension-1 braneworlds.

The above characteristics can be used to constrain the model in
question with respect to cosmological observations. Furthermore, on
the theoretical side it would be interesting to have particular bulk
solutions manifesting the cosmology evolution we have found, and in
particular setting the boundary conditions in order to fix $\beta$.
The most important information we have given here in this direction
is that it is pure six-dimensional Gauss-Bonnet gravity that gives the dynamical
LFRW-like evolution on the brane. The six-dimensional Einstein term gives the
possibility for a phase of geometric acceleration.
Furthermore, a varying $\beta$ could have interesting consequences
for a cosmological self-tuning scenario following equation
(\ref{betacon}). However, note that the evolution in $\beta$ would
then leave an imprint on the cosmological evolution equations. These
are amongst the open, interesting questions that this work puts
forward and which we hope will be answered in the near future.

\[ \]
{\bf Acknowlegements} We would like to thank Ruth Gregory for
numerous insightful comments and discussions on the subject of
codimension-2 defects. We  also thank N. D. Kaloper and R. Zegers for
critical discussions on codimension-2 braneworld cosmology and
junction conditions. C.C. acknowledges with great pleasure
early discussions with Brandon Carter on the self gravity of
topological defects. Brandon Carter's comments were not immediately
understood by both parties but were insightful and correct. C.C. would like to thank the
sunny ICG  in Portsmouth and the Department of Physics in Crete, and
in particular Elias Kiritsis, for their kind hospitality during
numerous stages of this work. A.P. would like to thank the Laboratoire de
Physique Th\'eorique in Orsay and the Department of Physics in Crete
for their kind hospitality during several stages of this work. G.K. is supported in part by the EU grants INTERREG IIIA
(Greece-Cyprus), MRTN-CT-2004-512914 and the
FP7-REGPOT-2008-1-CreteHEPCosmo-228644. A.P.
is supported by a Marie Curie Intra-European Fellowship EIF-039189.

\appendix

\section{The second order equations of motion} \label{2ndorder}

In this Appendix we will list the equations of motion which involve
the second order variables $A_2$, $N_2$ and $\b_2$. As we saw in the
main text, they depend on the matching conditions and also the
regulatisation that we have taken. For this Appendix, we
assume  that the bulk is empty, \ie ${\cal T}_{MN}=0$, and furthermore that the
three dimensional space is flat, \ie $\k=0$.

Let us first see for the topological matching conditions without an
explicit regularisation. The ${\cal O}(1/\ep)$ parts of $(tt)$ and
$(ij)$ equations and the   ${\cal O}(1)$ part of $(\th\th)$ equation
form an algebraic set of equations for the three second order
variables. We can write the system in the suggestive form \be D
\left( \begin{array}{ccc} A_2\\N_2 \\ \b_2 \end{array} \right) = C~
\label{ANbsystem}, \ee with the following coefficient matrix \be
D=\renewcommand{\arraystretch}{1.5}
\left(\begin{array}{ccc}2&0&{1 \over A \b}\left(H^2 + {1 \over 2\a}  -A^2\right) \\
2\left( 1 + {A \over N}   \right) & 2{A \over N}&-{1 \over N \b}\left(H^2 +{3 \over 2 \a}- A^2 -2 A N +  2{\ddot{a} \over a} \right)\\
{2 \a \over 3}\left( H^2 + {3 \over 2 \a} -A^2 -2 AN + 2 {\ddot{a} \over a}\right)&~~{2\a \over 3}\left( H^2 + {1 \over 2 \a}-A^2  \right)&0\end{array}\right)~\label{Dcoef},
\ee
and the following vector of the right hand side
\be
C=
\renewcommand{\arraystretch}{1.5}
\left(
\begin{array}{ccc} H^2 + {3 \over 2\a}  -A^2\\H^2 +{3 \over 2 \a} -3A^2 + 2 {A \over N} \left( {\ddot{a} \over a} + {3 \over 2 \a} \right)   \\   H^2 -A^2+ {\ddot{a} \over a} -AN + {2 \a \over 3}\left[ A^3 N + H^2 (2 A^2 +2N^2 - 5AN)  + 4 (A+N) \dot{A} H  + 2\dot{A}^2 +{\ddot{a} \over a}(A^2+H^2)\right]
\end{array}
\right)~. \label{Ccoef} \ee The solution of this system is given in
Sec.\ref{consistencysection}. There are two more equations that have
to be verified from the grouping of Fig.\ref{matrix}. First, the
${\cal O}(1)$ part of the $(rr)$ equation. This is again an
algebraic equation in the set of  $(\b_2, N_2, A_2)$ variables \bea
\left[1-2\a A (A+2\a N)+{2 \a \over 3}\left( H^2 + 2 {\ddot{a} \over a} \right)\right] A_2 + \left[ 1-6\a A^2 +2\a H^2 \right]N_2 ~~~~~~~~~~~~~~~~~~~~~~~~~~~~~~~~~~~~~~~~~~~~~~~~ \nn \\+ {A \over 3 \b}\left[ {3 \over 2} - \a A^2 -3 \a AN +{1 \over 2} {N \over A}+\a H^2 +\a {N \over A} H^2 +2\a {\ddot{a} \over a}    \right]\b_2 = \nn \\
=H^2 -AN +{1 \over 3}N^2 + {\ddot{a} \over a} + {1 \over 3}{\ddot{\b} \over \b} -2\a A^2 \left( A^2 +{5 \over 3}A N + N^2 \right)  + 2\a H^2 \left({1 \over 3} A^2  + {1 \over 3} AN + N^2 \right) ~~~~~~~~~~~~~~~~~~\nn \\
-4 \a H \left( A\dot{A}+ {1 \over 3}N \dot{A} - {1 \over 3}A \dot{N} \right) + {2 \a \over 3}{\ddot{a} \over a}\left( A^2 + 4AN +H^2 \right)
-{4 \a \over 3} A \ddot{A} +{2 \a \over 3}{\ddot{\b} \over \b}(H^2 - A^2)~.
\eea

On the other hand, the  ${\cal O}(1)$ part of the $(rt)$ equations
is a differential (Bianchi) equation with respect to the second
order variables \bea
4\a \dot{A_2} -{2 \a \over A \b} \left( H^2 + {1 \over 2 \a}  -A^2  \right)\dot{\b_2} + 4\a \left({\dot{A} \over A} + \left(2 - {N \over A}\right)H \right)A_2  - 4 \a H N_2 + {2\a A \over  \b} \left(H \left(1+  {N \over A}\right) +  {\dot{A} \over A}\right)\b_2= \nn \\
=3H\left(1 - {N \over A}\right) + 3{\dot{A} \over A} +2 \a H \left( H^2 \left(1- {N \over A} \right) + 3A^2- 2N^2 -AN \right) + 2 \a {\dot{A} \over A}(H^2 + 3A^2 +2 AN)~.
\eea

For the case of ring regularisation of the topological case, we have
as second order variables the set ($A_2$, $N_2$, $\b_2$,
$\bar{\b}_2$). In the exterior of the ring, the system of equations
in the same as above, with the simplification that $\dot{\b}=0$. In
the interior of the ring, by taking the simple relation
\reef{barb2}, we can trade $\bar{\b}_2$ for $\b_2$. Then, we can
easily see that we have again the same system \reef{ANbsystem}, with
\reef{Dcoef}, \reef{Ccoef}. On the other hand,  the ${\cal O}(1)$
part of the $(rr)$ equation changes to \bea
\left[1-2\a A (A+2\a N)+{2 \a \over 3}\left( H^2 + 2 {\ddot{a} \over a} \right)\right] A_2 + \left[ 1-6\a A^2 +2\a H^2 \right]N_2 ~~~~~~~~~~~~~~~~~~~~~~~~~~~~~~~~~~~~~~~~~~~~~~~~~~~~~~~ \nn \\+ {A \over 3 \b}\left[ {3 \over 2} - \a A^2 -3 \a AN +{1 \over 2} {N \over A}+\a H^2 +\a {N \over A} H^2 +2\a {\ddot{a} \over a}    \right]\b_2 = \nn \\
=\b H^2 - \b AN +{1 \over 3}N^2+ (1-\b)A^2 + \b {\ddot{a} \over a}  -2\a A^2 \left( A^2 + {(6-\b) \over 3}A N  + N^2 \right)   + 2\a H^2 \left({1 \over 3} A^2   + {(2-\b) \over 3} AN  + N^2 \right) \nn \\
-4 \a H \left( A\dot{A}+ {1 \over 3}N \dot{A} - {1 \over 3}A \dot{N} \right)
+ {2 \a \over 3}{\ddot{a} \over a}\left( (2-\b) A^2 + 4AN +\b H^2 \right)
-{4 \a \over 3} A \ddot{A} ~.~~~~~
\eea
Finally, the  ${\cal O}(1)$ part of the $(rt)$ equation also changes to
\bea
{4\a \over \b} \dot{A_2} -{2 \a \over A \b^2} \left( H^2 + {1 \over 2 \a}  -A^2  \right)\dot{\b_2} + {4\a \over \b} \left({\dot{A} \over A}  + \left(2 - {N \over A}\right)H \right)A_2  - {4 \a \over \b} H N_2 + {2\a A \over  \b^2} \left(H \left(1+  {N \over A}\right) +  {\dot{A} \over A}\right)\b_2= \nn \\
=3H\left(1 - {N \over A}\right) + 3{\dot{A} \over A} +2 \a H \left( H^2 \left(1- {N \over A} \right) +\left({4 \over \b}-1\right) A^2- {2 \over \b}N^2 -\left({2 \over \b}-1\right)AN \right) \nn\\
+ 2 \a {\dot{A} \over A}(H^2 + \left({4 \over \b}-1\right)A^2 +{2 \over \b} AN)~.~~~~~~~~~~~~~~~~~~
\eea

\section{Comparison with the double Wick rotated black hole solution} \label{bh}

In this Appendix we will compare the explicit simple solution of a
system where the bulk has a Gauss-Bonnet term and cosmological
constant, with the constraint that we noted in sec. \ref{vanishing}.
We will show that firstly, in this solution it is $K'_{\m\n} \neq
0$, violating the assumptions of \cite{bos} and furthermore that the
leading terms in the $r$ expansion of $K_{\m\n}$ is the one found in
the main text from the local consistency conditions.

The solution of a Gauss-Bonnet $AdS$ black hole with horizon of
toroidal topology is
\bea
ds^2&=&-V(R)dt^2 + { dR^2 \over V(R)} +
R^2 \d_{mn} dx^m dx^n  ~, \\
V(R)&=&{R^2 \over 2\a} \left[ 1- \sqrt{1- 4\a \left( k^2 - {\m \over
R^5 }\right)} \right]  ~, \eea with $k^2=-\La_6/10$. Let us assume
that $\a>0$, $\m>0$, $4\a k^2<1$. Then, there is only one
singularity at $r=0$, shielded by an horizon at
$R_H=(\m/k^2)^{1/5}$.

Let us now make a double Wick rotation $t \to i \th$ and $x^0 \to i
t$. Then, we obtain the metric \be ds^2=R^2 \eta_{\m\n} dx^\m dx^\n
+ { dR^2 \over V(R)} +V(R)d\th^2 ~. \ee We can go to a
Gaussian-Normal system by means of the transformation $R' =
\sqrt{V}$ and obtain the metric \be
 ds^2=R(r)^2 \eta_{\m\n} dx^\m
dx^\n + dr^2 +V(R(r))d\th^2 ~.
\ee
Then, the comparison with the
previous presentation is straightforward
\bea
&&L=\sqrt{V} ~, \\
&&K_{\m\n}= {1 \over 2}(R^2)'\eta_{\m\n}  =  R \sqrt{V} \eta_{\m\n} ~,
\eea
with the requirement that $R_H=1$ (thus $\m=k^2$) in order to have
the induced metric $g_{\m\n}|_{R_H}=\eta_{\m\n}$. The coordinate
transformation can be written locally as
\be
R=1+ {5 \over 4}k^2
r^2  + {\cal O}(r^4) ~.
\ee
Therefore, the above quantities are
\bea
&&L={5 \over 2}k^2 r + {\cal O}(r^3) \label{bhL} ~,\\
&&K_{\m\n}= {5 \over 2} k^2 r \eta_{\m\n} + {\cal O}(r^2) ~, \eea
from which we confirm that $\b_2=0$ and verify also the correctness
of \reef{bhtest}. This is because
$K'_{\m\n}|_{r=0}=(-N_2,A_2,A_2,A_2)$, with $N_2=A_2$ because of the
Minkowski symmetry. Then, the relation \reef{bhtest} gives \be
A_2=N_2=-{\La_6 \over 4}~. \ee

Moreover, from  \reef{bhL} we obtain $\b$, in the case that $\th$ is
normalized so that $\th \in [0,2\pi)$.


\end{document}